# Multi-channel electrically tunable varifocal metalens with compact multilayer polarization-dependent metasurfaces and liquid crystals

Zhiyao Ma, Zhe Li, Tian Tian, Yuxuan Liao, Xue Feng*, Yongzhuo Li, Kaiyu Cui, Fang Liu, Hao Sun, Wei Zhang and Yidong Huang*

Department of Electronic Engineering, Tsinghua University, Beijing 100084, China

*Corresponding author: x-feng@tsinghua.edu.cn (X.F.); yidonghuang@tsinghua.edu.cn (Y.H.)

**Abstract**

As an essential module of optical systems, varifocal lens usually consists of multiple mechanically moving lenses along the optical axis. The recent development of metasurfaces with tunable functionalities holds the promise of miniaturizing varifocal lens. However, existing varifocal metalenses are hard to combine electrical tunability with scalable number and range of focal lengths, thus limiting the practical applications. Our previous work shows that the electrically tunable channels could be increased to $2^N$ by cascading $N$ polarization-dependent metasurfaces with liquid crystals (LCs). Here, we demonstrated a compact eight-channel electrically tunable varifocal metalens with three single-layer polarization-multiplexed bi-focal metalens and three LC cells. The total thickness of the device is ~6 mm, while the focal lengths could be switched among eight values within the range of 3.6 to 9.6 mm. The scheme is scalable in number and range of focal lengths and readily for further miniaturization. We believe that our proposal would open new possibilities of miniaturized imaging systems, AR/VR displays, LiDAR, *etc*.

# Introduction

Varifocal lens is an essential component in optical imaging and display systems. Usually, the required phase gradient for a lens is achieved by lightwave propagation through a curved bulky transparent material, while multiple mechanically moving lenses along the optical axis are required to vary the focal length. Consequently, it is a great challenge to miniaturize the varifocal lens for practical applications. One possible approach to address such challenge is to employ the metasurface. Metasurface is developed by arranging subwavelength scatters in a plane for controlling the property of lightwave, including phase[1,2], amplitude[3–5], polarization[6,7], frequency[8,9], *etc*. Metasurface lens (also called metalens)[10–14] could achieve the phase gradient of a lens within subwavelength thickness and resolution, thus possessing higher compactness, higher N.A. (numerical aperture)[15,16] and even richer functionalities[17–19] compared with the traditional lens. Instead of mechanically moving multiple layers along the optical axis, the focal length of metalenses could be tuned by various mechanics, including reconfigurable metasurface[20–23], mechanically stretching[24–27], cascaded Moiré metalens[28–30], polarization multiplexing[31,32], OAM (orbital angular momentum) multiplexing[33,34], *etc*.

However, pixel-by-pixel active reconfigurable metalenses are mainly operating in microwave regime[22], since reconfiguring optical subwavelength structures would face significant technical difficulty. Other reconfiguring mechanics like phase change materials[20,23] and thermo-optical effects[20,21] are usually limited in number or range of focal lengths. Mechanically stretching metalens[24–27] and Moiré metalens[28–30] still rely on mechanically moving, which limits the speed and lifetime of the device. For approaches of switching the focal length by controlling certain properties of input light, *e.g.* polarization[31,32] and carried OAM[33], the number of focal lengths is also limited. Polarization multiplexing could only achieve two focal lengths due to the number of states[31,32]. while the channels of OAM multiplexed metasurface is still less than five to meet the required information density[33,34]. For an electrically tunable varifocal optical metalens with scalable number of focal lengths, one solution is exploiting the DoFs of multiple cascaded polarization-dependent metasurfaces. Specifically, we have recently proposed an $N$-layer cascaded structure of polarization-dependent metasurfaces and LCs for $2^N$ electrically switchable channels of vortex beam generation and beam steering[35].

In this work, our previous proposal is extended to a compact $2^N$-channel electrically tunable varifocal metalens with $N$ single-layer polarization-multiplexed bi-focal metalenses and $N$ LC cells. By controlling the input polarization states through a LC cell, the focal-length of each single-layer metalens could be switched between two values. Then, the equivalent focal-length of a $N$-layer cascaded metalens could be switched in $2^N$ values, corresponding to the $2^N$ combinations of input polarization states at each single-layer metalens. The number, resolution and range of focal lengths would be scalable with increased number of layers. Experimentally, by alternately stacking three single-layer metalenses and three LC cells, we have demonstrated a varifocal metalens with thickness of ~6 mm and eight electrically switchable channels. The focal lengths could be switched among eight values within the range of 3.6 to 9.6 mm. The focused FWHM (full-width at half-maximum) is within the range of 19.1 to 40.6 μm, while the efficiency is within the range of 5.4% to 10.3%.

# Results

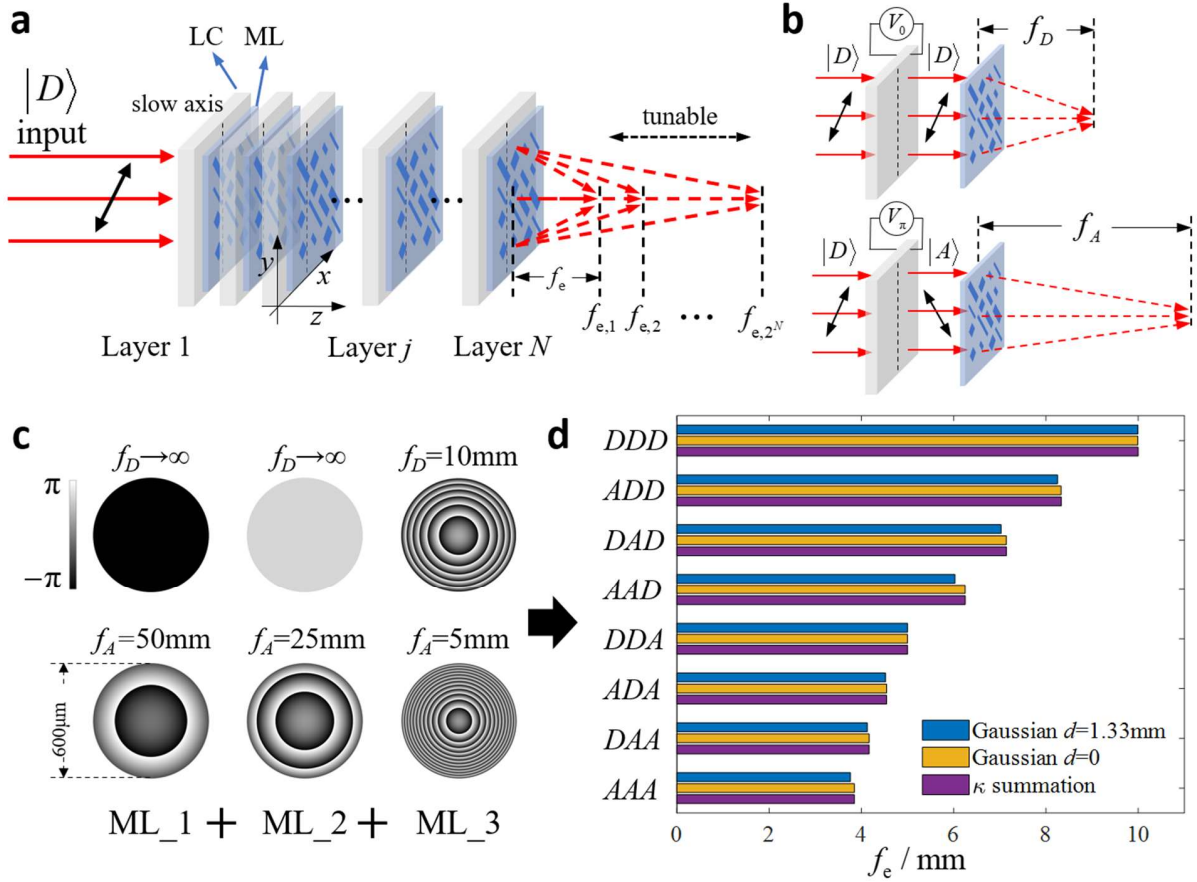

Fig. 1 **Principle and design of compact metalens**. (**a**) Schematic of compact metalens with alternately cascaded $N$ single-layer metalenses (ML) and $N$ LC cells. (**b**) One layer of polarization-multiplexed bifocal metalens with a LC cell attached on the input side, so that the focal length could be switched between $f_D$ and $f_A$ according to the input states $|D\rangle$ and $|A\rangle$. (**c**) Phase profiles for $|D\rangle$ and $|A\rangle$ input of each single-layer metalens with the corresponding focal lengths. (**d**) Calculated whole focal lengths corresponding to eight input polarization combinations. Results by Gaussian beam propagation with $L_{exp}$ = 1.33 mm, $L$ = 0, and phase profile summation are compared.

Fig. 1(a) is the schematic of compact metalens with alternately cascaded $N$ layers, while each layer consists of a metasurface and a LC cell attached on the input side as shown in Fig. 1(b). The metasurface is designed as a polarization-multiplexed bifocal metalens, where two lens phase profiles with different focal lengths could be modulated on the input lightwave with two orthogonal polarization states, respectively[31,32]. Here, the orthogonal input states at each bifocal metalens are designed as diagonal and anti-diagonal linear polarization (denoted as $|D\rangle$ and $|A\rangle$), which could be manipulated by the vertical slow axis of LC. Specifically, the input state would be varied from one to the other with the π phase retardance, while it would keep constant with 0 phase retardance. Fig. 1(b) shows the case of $|D\rangle$ input without loss of generality. Thus, the focal lengths of each single-layer metalens could be independently switched between $f_D$ and $f_A$.

After cascading $N$ such layers, the total combinations of the input polarization state at each single-layer metalens would be $2^N$. Supposing that all metalenses were attached with sufficiently close distance, the functionality of the whole cascaded structure could be expressed as the summation of the phase profile loaded on each single-layer metalens according to the input state. Here, the phase profile of each single lens is expressed in spatial coordinates $(x,y)$ as Eq. 1[36], where $\lambda$ is the wavelength and $f$ is the focal length. Eq. 1 is a linear expression of $1/f$, thus the summation of multiple profiles is still a lens profile with different focal length as shown in Eq. 2, which is a necessary condition for the following design of $2^N$ channels[35].

$$\varphi(x,y,f) = \frac{\pi(x^2+y^2)}{\lambda f} \tag{1}$$

$$\varphi_{\boldsymbol{\mu}} = \sum_{j=1}^{N} \varphi_{j,\mu_j} = \sum_{j=1}^{N} \frac{\pi(x^2+y^2)}{\lambda f_{j,\mu_j}} = \frac{\pi(x^2+y^2)}{\lambda} \sum_{j=1}^{N} \frac{1}{f_{j,\mu_j}} \tag{2}$$

$$f_{e,\boldsymbol{\mu}} = \sum_{j=1}^{N} \frac{1}{f_{j,\mu_j}} \tag{3}$$

In Eq. 2, $\boldsymbol{\mu}$ is a $N \times 1$ vector representing the input polarization state at each single-layer metalens. It can be seen that the functionality of the whole structure is still a lens and the equivalent focal length of whole cascading structure $f_e$ could be expressed in summation of reciprocal forms as Eq. 3. Thus, the equivalent focal lengths can take $2^N$ distinct value according to $2^N$ values of $\boldsymbol{\mu}$. It should be noted that Eq. 1 is chosen as the parabolic phase profile instead of the commonly used hyperbolic phase profile (see Supplementary materials)[10–13] to meet the condition that the phase profile is a linear expression of certain parameters. Actually, Parabolic lens profile is almost equivalent to the commonly used hyperbolic phase profile for current demonstration, since the N.A. is low and only normal incidence is considered. While for more general cases, although parabolic phase profile would introduce spherical abbreviation, it has been proven to exhibit wider field-of-view (FOV) for imaging compared with hyperbolic phase profile[37].

Moreover, the thickness of substrates and liquid crystals have to be considered for a practical implementation. Instead of simply summing the phase profiles of single-layer metalenses, the theoretical modelling of the whole structure should include lightwave propagation between adjacent metalenses. Here, since paraxial approximation could be satisfied in the demonstration, $q$ parameters of Gaussian beam are utilized to model the cascaded lens profiles with proper distance[38]. Suppose the $q$ parameter of input Gaussian beam is $q_{in}$, while the optical distance between adjacent phase profiles is $d$. As property of $q$ parameters, the propagation effects through a lens or free space can be both considered as applying four parameters $T_{11}$, $T_{12}$, $T_{21}$, $T_{22}$ on $q$ as the expression of Eq. 4. The values of the four parameters are determined by the focal length $f$ or distance $L$, which could be written in the matrix form of $\mathbf{T}_{lens}(f)$ or $\mathbf{T}_{space}(L)$ in Eq. 5. Then, the propagation effects of multiple cascaded layers could be considered as applying the corresponding matrices on $q$ in sequence, which are exactly equivalent to applying the multiplication of the matrices on $q$. Thus, the $q$ parameter at the output plane (denoted $q_{out}$) could be expressed in a cascaded matrix multiplication form as Eq. 6, where the output plane is regarded as the top surface of the $N$-th single-layer metalens. Finally, the equivalent focal length $f_e$ of the whole cascaded metalens is defined as the distance between the output surface and the waist of the output beam, which could be calculated with Eq. 7. It should be mentioned that the whole cascaded metalens with distance between adjacent single-layer metalenses is not equivalent to a single lens, and the definition of focal length is similar to the back focal length of an imaging system.

$$q' = \frac{T_{11}q_{in} + T_{12}}{T_{21}q_{in} + T_{22}} \quad (4)$$

$$\mathbf{T}_{lens}(f) = \begin{pmatrix} 1 & 0 \\ -1/f & 1 \end{pmatrix}, \mathbf{T}_{space}(L) = \begin{pmatrix} 1 & L \\ 0 & 1 \end{pmatrix} \quad (5)$$

$$\mathbf{T}_{whole} = \prod_{j=1}^{N} \mathbf{T}_{space}(L)\mathbf{T}_{lens}(f_{j,\mu_j}) \quad (6)$$

$$f_e = -\mathrm{Re}(q_{out}) \quad (7)$$

With the Gaussian beam propagation model, the parameters of a three-layer cascaded metalens are designed and shown in Fig. 1(c) and (d). Fig. 1(c) shows the phase profiles for $|D\rangle$ and $|A\rangle$ input of each single-layer metalens with the corresponding focal lengths. It should be mentioned that a lens with $f\to\infty$ corresponds to a uniform phase profile, which actually carry out no operation on the wavefront. Fig. 1(d) shows the calculated whole focal lengths corresponding to eight input polarization combinations of each single-layer metalens. For example, *DAA* represents that the input state at the first single-layer metalens is $|D\rangle$, while the input state at the second and third single-layer metalens are $|A\rangle$. The equivalent propagation distance between two adjacent single-layer metalenses is $L_{exp}$ = 1.33 mm, which is obtained by the experiments (see Eq. 8). Then, the calculated results by Gaussian beam propagation (Eq. 4 ~ Eq. 7) with $L_{exp}$ = 1.33 mm and $L$ = 0, as well as phase profile summation (Eq. 3) are compared. It can be seen that the propagation results with $L_{exp}$ = 1.33 mm differs from summation results, while the propagation results with $L$ = 0 are almost consistent with summation results. The Gaussian beam propagation model with $L$ = 0 reduces to phase profile summation correctly. Detailed derivations, numerical results and related discussions could be found in Supplementary materials.

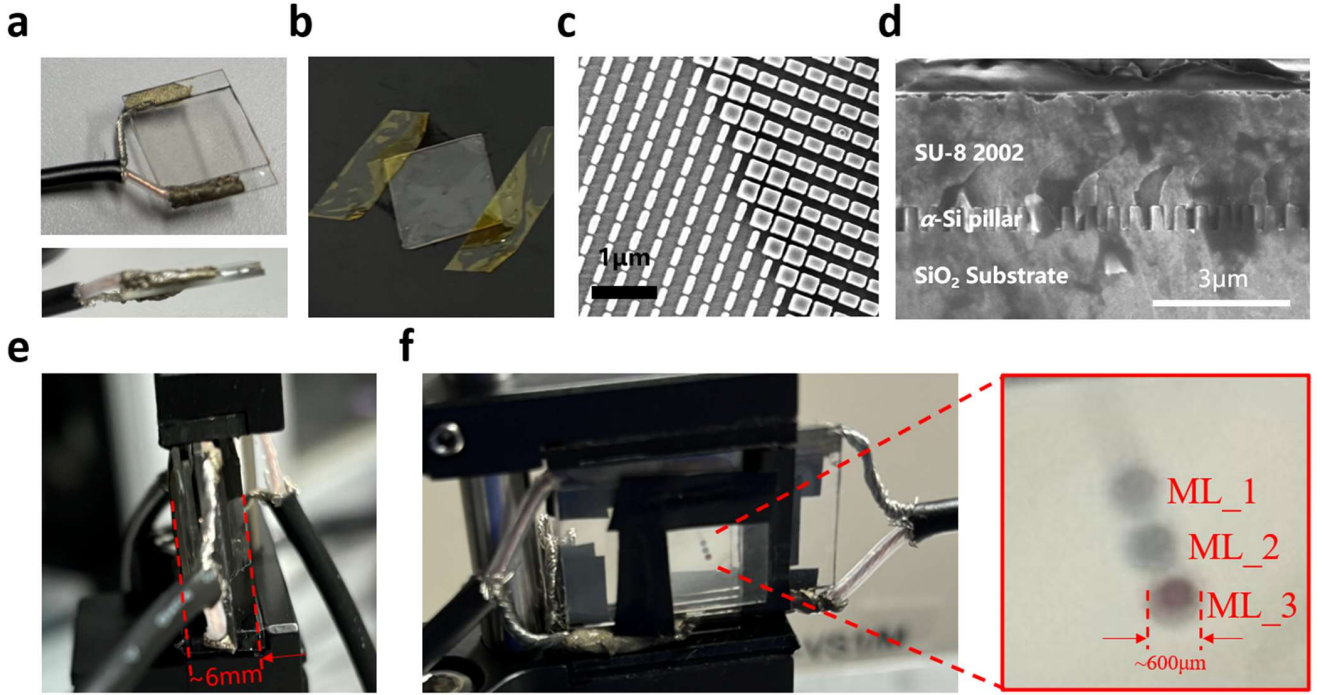

Fig. 2 **Experimental implementations.** (**a**) Photograph of LC cells customized from JCOPTIX, total size 2 cm × 2 cm × 1.45 mm. (**b**) Photograph of single-layer metalens with SU-8 spin-coated, effective diameter 600 μm, total size 1 cm × 1 cm × 0.54 mm. (**c**) SEM image of single-layer metalens without SU-8. (**d**) SEM image of cross section of single-layer metalens with SU-8. (**e**)(**f**) photographs of the implemented three-layer metalens. (**e**) Side view. (**f**) Normal view. Inset is close view of three single-layer metalenses.

To implement the cascaded metalens, LC cells and single-layer metalenses are prepared at first. The LC cells are customized from JCOPTIX with photograph shown in Fig. 2(a). The thickness of one LC cell is $t_{LC}$~1.45 mm in total, which mainly consists of the two glass substrates since the thickness of LC layer is only ~10 μm, while the size is 2 cm × 2 cm. Besides, the voltages corresponding to 0 and π phase retardance are calibrated, respectively. The single-layer metalenses are designed and fabricated $a$-Si nanopillars with height of 500 nm. The photograph and SEM image of fabricated metalens sample without spin-coating are shown in Fig. 2(b) and Fig. 2(c), respectively. To protect the nanopillars, SU-8 2002 photoresist is spin-coated as a spacer layer. Fig. 2(d) shows SEM image of the cross section of the spin-coated single-layer metalens. It can be seen that the SU-8 (thickness of ~3 μm) fills all the gaps of the nanopillar (height of ~500 nm). Thus, the total thickness $t_{ML}$~0.54 mm of one single-layer metalens mainly consists of the $SiO_2$ substrate as well. Besides, the diameter of each metalens is 600 μm, while the total size of the substrate is 1 cm × 1 cm. From the experimental parameters, it can be seen that paraxial approximation could be satisfied since all of the focal-lengths are more than 3 mm while the diameter of each metalens is 600 μm. Also, propagation through a medium of thickness $t$ and refractive index $n$ is equivalent to that through a free space of distance $L=t/n$[39], so that the equivalent propagation distance between two adjacent single-layer metalenses could be obtained by Eq. 7 with $n_{SiO_2}$ ~1.5. Details of LC calibration, metasurface design and fabrication are presented in the section of Methods and Supplementary materials.

$$L_{exp} = (t_{LC} + t_{ML})/n_{SiO_2} \sim 1.33 \text{ mm} \qquad (8)$$

Then, the cascaded metalens is implemented by alternately stacking three LC cells and three single-layer metalenses with parameters shown in Fig. 1(b). To start with, since the transverse size of LC cells is larger than that of metalens substrates, the first metalens is attached on the right area of the first LC cell. While the second LC cell is attached on the first metalens with a large displacement to the right relative to the first LC cell, so that the tape attaching and wire connecting would be convenient. Next, the second metalens is attached on the second LC cell and roughly overlaps with the first metalens. For more precise alignment, the marks on both metalenses are imaged to qualitatively observe the relative displacement, and the attached location of the second metalens could be slightly adjusted until the marks align well. After that, the third LC cell is attached on the second metalens and overlaps with the first LC cell. Finally, the attaching and alignment of the third metalens follow that of the second metalens. For the implemented sample, the photograph of the side view and normal view are shown in Fig. 2(e) and (f), respectively. As seen in Fig. 2(e), the total thickness of the cascaded metalens is ~6 mm, which agrees with the size summation of all devices. The inset of Fig. 2(f) clearly shows the effective regions of the three single-layer metalenses with diameter of ~600 μm, whose centers are aligned along a line. The final alignment error in $x$-$y$ plane is less than 34 μm. Details of the alignment process and discussion for future improvement could be found in Supplementary materials.

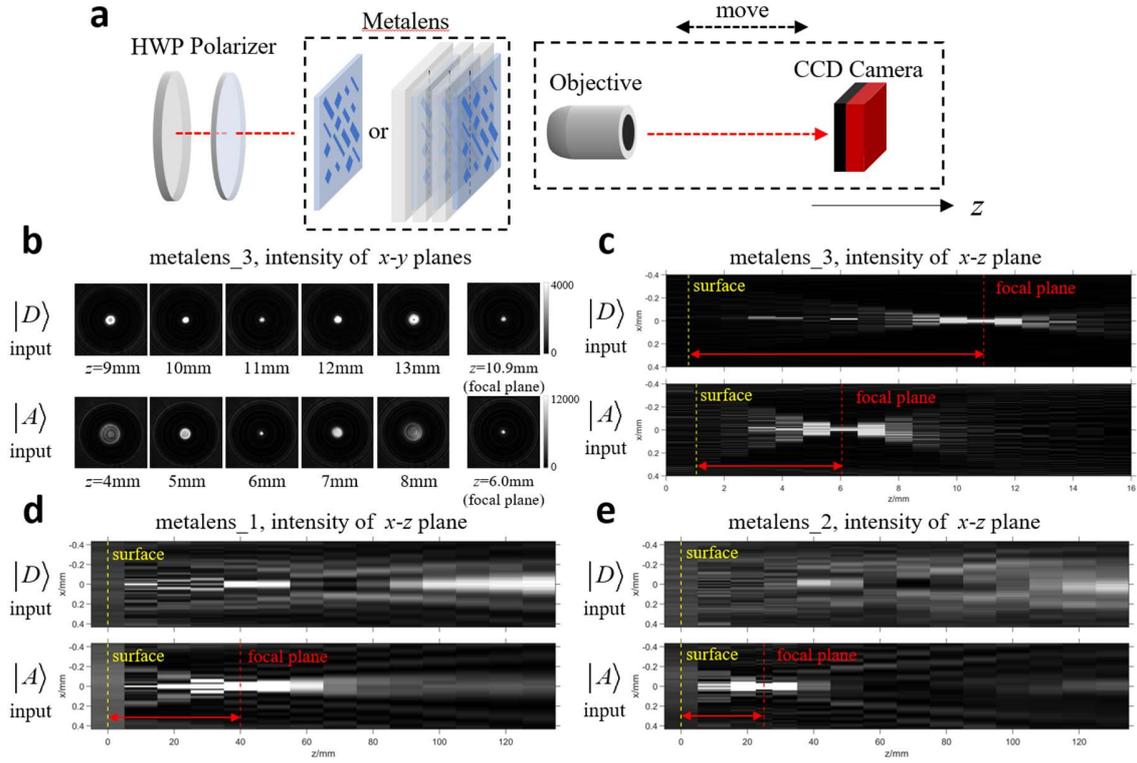

Fig. 3 **Single metalens measurement.** (**a**) Schematic of optical setup for single and cascaded metalens measurement. (**b**) Measured output intensity profiles at *x-y* planes around and at the focal plane of metalens 3. (**c**) Measured intensity profiles at *x-z* plane of metalens 3. (**d**) Measured intensity profiles at *x-z* plane of metalens 1. (**e**) Measured intensity profiles at *x-z* plane of metalens 2.

For experimental characterization, the focal length of all three single-layer metalenses and the cascaded one are measured. The schematic of the optical setup is show in Fig. 3(a). The input Gaussian beam could be filtered to $|D\rangle$ or $|A\rangle$ polarization, while the whole imaging system could move along *z* axis to measure the displacement quantitatively. Thus, the output field at difference *x-y* planes along *z* axis could be measured and captured by the CCD camera. Among the intensity profiles of output field, the output surface could be identified by the mark on the substrate of the third metalens, while the focal plane could be identified by optimal focal point. The focal length could be acquired by the *z* position difference between the output surface and the focal plane of the output beam.

For single-layer metalenses, the input beam is switched between $|D\rangle$ and *A* polarization. Since the focal lengths of the third single-layer metalens are relatively short, it is measured with 1.00 mm step of *z*, and the focal planes are identified particularly. The results are shown in Fig. 3(b) and (c). Fig. 3(b) is the output intensity profiles at *x-y* planes with several *z* positions around and at the focal plane, while Fig. 3(c) is the replotted intensity profiles at *x-z* plane with *y* position at the center of the metalens. The measured focal lengths with $|D\rangle$ and $|A\rangle$ input are 10.16 mm and 5.00 mm respectively, agreeing with the designed values of 10 mm and 5 mm, respectively. Since the focal lengths of the first and second single-layer metalens are relatively long, the step of *z* is chosen as 10 mm without identifying the focal planes. Similarly, the replotted intensity profiles at *x-z* plane of the first single-layer metalens are shown in Fig. 3(d). The output beams with $|D\rangle$ input is not apparently focused, agreeing with the designed focal length of $f \rightarrow \infty$. The measured focal length with $|A\rangle$ input is 40~50 mm, whose error fron designed 50 mm is within the 10 mm step of measurement. Also, the replotted intensity profiles at *x-z* plane of the second single-layer metalens are shown in Fig. 3(e). The measured focal lengths of ~∞ and ~25 mm agree with the original design.

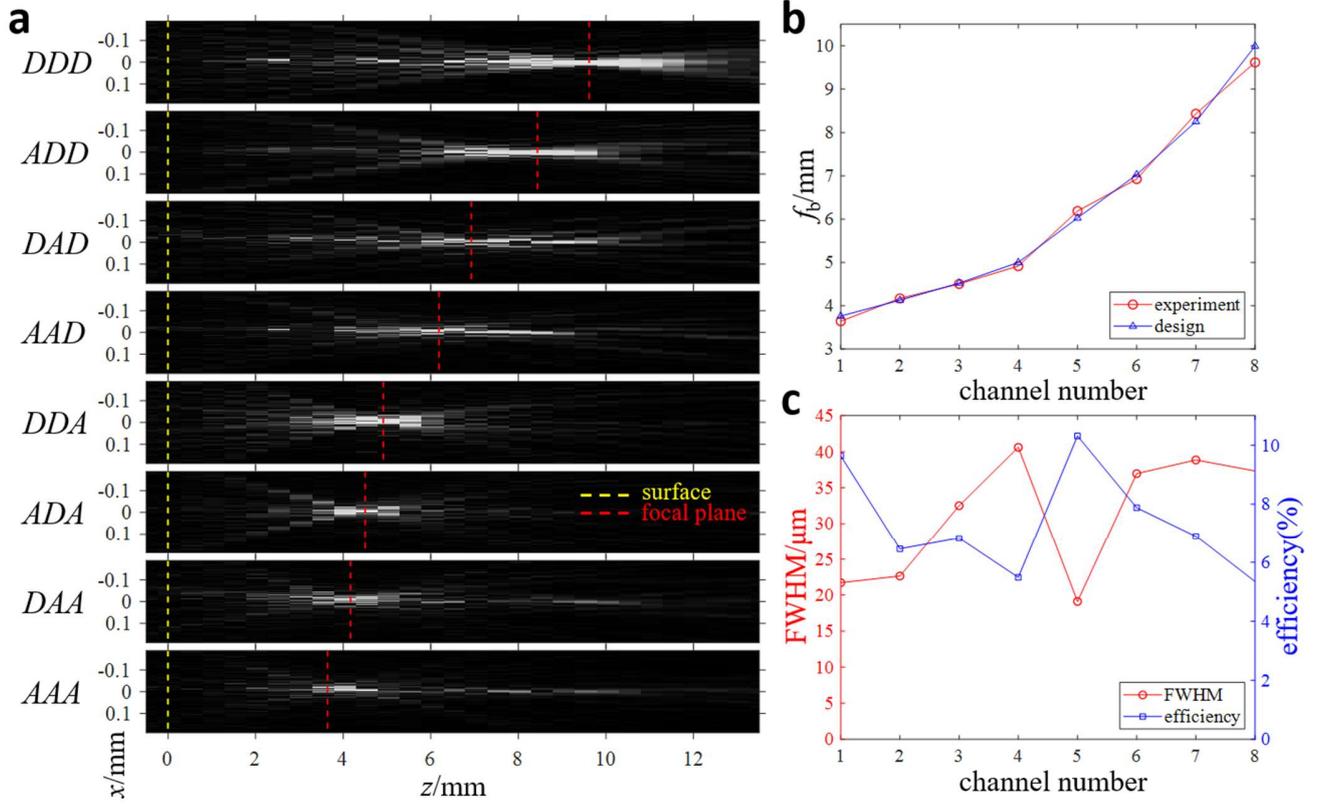

Fig. 4 **Cascaded metalens measurement.** (**a**) Intensity profiles at *x-z* plane of eight channels. The surface and focal plane are marked by yellow and red dash lines, respectively. (**b**) Comparison of measured focal lengths and designed focal lengths. (**c**) Measured FWHM and efficiency of eight channels.

For the whole cascaded metalens, the input beam is fixed on $|D\rangle$ since the polarization states are switched by LCs. The step of *z* is also 1.00 mm and the focal planes are identified particularly. The replotted intensity profiles at *x-z* plane of all eight channels are shown in Fig. 4(a). It can be seen from Fig. 4(a) that the measured intensity profiles and corresponding focal lengths of eight channels are distinct from each other. Then, as shown in Fig. 4(b), the trend of the measured focal lengths agrees with the designed values, where the errors are less than 3.8%. Therefore, the focal length of the whole cascaded metalens could be switched among the measured eight values within the range of 3.6 to 9.6 mm. In addition, the measured FWHM and efficiency is plotted in Fig. 4(c). The FWHM at the focal plane is measured by 2-dimensional Gaussian fitting, which is within the range of 19.1 to 40.6 μm. While the efficiency is defined the ratio of the total intensity around the focal spot to the intensity of the input, which is within the range of 5.4% to 10.3%. Details of the measurement and data processing could be found in Supplementary materials.

## Discussion

Compared with reported varifocal metalens, our demonstration could achieve scalable number and range of focal lengths by electrically tuning. The range of focal lengths is determined by that of a single-layer bi-focal metalens, while the number resolution of focal lengths would increase with the number of cascaded layers. A comprehensive comparison of reported varifocal metalens is presented in Supplementary materials.

The demonstrated FWHM and efficiency could be further optimized in various ways. Firstly, the alignment of cascaded structure could be performed under a mounting system[40] to achieve subwavelength precision. Secondly, the reflection at multiple substrates of LC cells and metasurfaces could be further reduced by anti-reflection coating. Thirdly, the etching and spin-coating process of metasurface could also be improved to reduce fabrication error. What is more, most reported technics of high-performance single-layer metalens could be utilized in the cascaded metalens, including Pancharatnam-Berry phase[11–14], elimination of chromatic aberration[11–13], inverse design[41,42], *etc*.

Besides, to achieve more stacked layers and further miniaturization in practical applications, the thickness could be further reduced by improving the fabrication methods. The current thickness per layer is ~2 mm, which is mainly composed of the substrates of LC cell and metasurface. According to a previous report, the total thickness of the LC cell could be reduced to about 50 μm[43]. Furthermore, LCs could be packaged on the substrate of metasurfaces for a fully integrated scheme[44]. Reduced thickness is also crucial to leverage the wide FOV of the parabolic profile (see Eq. 1), since the alignment could be better preserved at obliquely incident.

To conclude, a compact eight-channel electrically tunable varifocal metalens are demonstrated by alternately stacking three single-layer polarization-multiplexed bi-focal metalens and three LC cells. The total thickness is ~6 mm, while the focal lengths are switchable among eight values within the range of 3.64 to 9.62 mm. Our scheme is scalable in resolution and range of focal lengths, as well as extensible for other operating wavelength, materials and design technics. Potential applications include miniaturized imaging systems, AR/VR displays, LiDAR, *etc*.

# Methods

**Calibration of liquid crystal**

The phase retardance of LC varies spatially since the thickness of LC layer is not uniform. Thus, the the voltages corresponding to 0 and π phase retardance should be calibrated within the corresponding region of input beam. Experimentally, the calibration of LC is conducted around the corresponding region of metalens, where the calibration optical setup is presented in Supplementary materials.

**Numerical Simulation**

The amplitude and phase modulation of rectangular nanopillars with different height, period, length and width are numerically calculated by the Finite Difference Time Domain (FDTD) method. The simulated nanopillar is coated with SU-8 photoresist. For the eventually used set, the height is 500 nm while the lattice constant is 400 nm, and the length and width are within the range of 80 nm to 320 nm. The detailed simulation and design process of the metalens can be found in Supplementary materials.

**Fabrication**

First, 500nm-thick $\alpha$-Si is deposited on a quartz substrate by PECVD. Then, a layer of Cr as metal hard mask are deposited by electron beam (EB) evaporation, and a layer of $SiO_2$ is grown on the Cr layer as an additional hard mask to avoid experimentally uncontrollable lift-off process of Cr. After that, the rectangular patterns are fabricated by electron beam lithography (EBL) and inductively coupled plasma reactive ion etching (ICP-RIE). The detailed fabrication process without spin-coating are similar to that of our previous work[45]. Subsequently, the sample is spin-coated with SU-8 2002 photoresist at 2000 rpm. Photo-lithography without mask and baking are performed to make a hard spacer layer.